\documentclass[]{aa}

\usepackage{graphicx}
\usepackage{txfonts}
\usepackage{xspace}
\usepackage{color}
\usepackage{url}
\usepackage{hyperref}
\usepackage[flushleft]{threeparttable}
\newcommand {\bc}{\begin {center}}
\newcommand {\ec}{\end {center}}
\newcommand {\be}{\begin {equation}}
\newcommand {\ee}{\end {equation}}
\newcommand {\beq}{\begin {eqnarray}}
\newcommand {\eeq}{\end {eqnarray}}

\def\flux{erg s$^{-1}$ cm$^{-2}$}
\def\lum{erg s$^{-1}$}

\def\nustar{{\it NuSTAR}\xspace}

\def\a05{A\,0535$+$262}
\def\maxi{MAXI~J0903$-$531\xspace}

\newcommand{\ero}{{eROSITA}\xspace}
\newcommand{\art}{{ART-XC}\xspace}

\newcommand{\srg}{{SRG}\xspace}

\begin{document}

\title{SRG/ART-XC, \textit{Swift}, NICER and \textit{NuSTAR} study of different states of the transient X-ray pulsar MAXI~J0903$-$531}

\author{Sergey S. Tsygankov \inst{1,2}
   		  \and Sergey V. Molkov \inst{1}
   		  \and Victor Doroshenko \inst{3,1}
   		  \and Alexander A. Mushtukov \inst{4,1}
   		  \and Ilya A. Mereminskiy\inst{1}
   		  \and Andrei N. Semena\inst{1}
   		  \and Philipp Thalhammer\inst{5}
   		  \and J\"orn Wilms\inst{5}
   		  \and Alexander A. Lutovinov \inst{1}
          }
          
   \institute{Space Research Institute of the Russian Academy of Sciences, Profsoyuznaya Str. 84/32, Moscow 117997, Russia    
       \and
              Department of Physics and Astronomy, FI-20014 University of Turku,  Finland \\ \email{sergey.tsygankov@utu.fi}
       \and
              Institut f\"ur Astronomie und Astrophysik, Universit\"at T\"ubingen, Sand 1, D-72076 T\"ubingen, Germany
       \and 
            Leiden Observatory, Leiden University, NL-2300RA Leiden, The Netherlands
            \and Dr.\ Remeis-Observatory \& Erlangen Centre for Astroparticle Physics, Friedrich-Alexander-Universit\"at Erlangen-N\"urnberg, Sternwartstr. 7, 96049 Bamberg, Germany
          }
          
\titlerunning{Different states of XRP MAXI~J0903$-$531}
\authorrunning{S.~Tsygankov et al.}

\date{2021}

\abstract{
The results of the broadband spectral and timing study of the recently discovered transient X-ray pulsar \maxi in a wide range of luminosities differing by a factor of $\sim30$ are reported. The observed X-ray spectrum in both states can be described as a classical pulsar-like spectrum consisting of the power-law with the high-energy cutoff. We argue that absence of the spectrum transformation to the two-hump structure expected at low fluxes points to a relatively weak magnetic field of the neutron star below $(2\mbox{--}3)\times10^{12}$~G. This estimate is consistent with other indirect constraints and non-detection of any absorption features which can be interpreted as a cyclotron absorption line. Timing analysis of the \nustar\ data revealed only slight variations of a single-peaked pulse profile of the source as a function of the energy band and mass accretion rate. In both intensity states the pulsed fraction increases from 40\% to roughly 80\% with the energy. Finally we were also able to obtain the orbital solution for the binary system using data from the {\it Fermi}/GBM, NICER and {\it NuSTAR} instruments.
}


\maketitle
%
\section{Introduction}
\label{intro}

The transient X-ray source \maxi was discovered with the MAXI/GSC nova alert system on April 14, 2021 with an indication of the gradual increase of the flux \citep{2021ATel14555....1Y}. An accurate source localisation with coordinates of $\alpha_\mathrm{J2000.0}=136\fdg27866$, $\delta_\mathrm{J2000.0}=-53\fdg50518$ ($2\farcs8$ uncertainty radius at the 90\% confidence level) was obtained two days later with the {\it Swift}/XRT telescope \citep{2021ATel14557....1K}. The same authors pointed to the possible optical counterpart, {\it Gaia} EDR3 source 5311384333263075840, with the distance estimated at around 10\,kpc \citep{2021AJ....161..147B}. A subsequent discovery of coherent pulsations with the period of $P=14.055$~s \citep{2021ATel14559....1R} by the NICER telescope allowed then to unambiguously identify the object as a transient X-ray pulsar (XRP). A spectral type of the companion star B1.5--2 III-Ve was determined using observations at the Southern African Large Telescope \citep[SALT;][]{2021ATel14564....1M}, suggesting that \maxi is a new Galactic Be X-ray binary system (Be/XRP).

Analysis of the long-term monitoring of the companion star from the ASAS-SN project revealed a strong modulation at a period of 3.84\,days with an amplitude of 0.05 magnitudes \citep{2021ATel14564....1M}. Another periodicity at 0.79\,days was found using the Transiting Exoplanet Survey Satellite (TESS) data, that could be interpreted as a pulsation period or a rotation period of a Be star \citep{2021ATel14568....1R}. The presence of this periodicity was later confirmed using the OGLE survey data \citep{2021ATel14569....1U} although the authors suggested the 3.8\,d periodicity to be a 1-day alias of 0.79\,d period. Later, based on the analysis of the long-term monitoring data from the {\it Swift}/BAT transient monitor and {\it Fermi}/GBM monitor another longer periodicity with period of around 57\,days was found by \cite{2021ATel14730....1C}. 
Assuming that this is the orbital period, the location of \maxi on the Corbet diagram \citep{1986MNRAS.220.1047C} supports its Be/XRP classification.

Based on the NICER data it was shown that the 0.5--10\,keV energy spectrum can not be desribed by just an absorbed power-law model, but an additional blackbody component with temperature of $kT=0.61\pm0.02$\,keV is required \citep{2021ATel14559....1R}. Closer to the end of the outburst, when \maxi faded to the 3--50\,keV flux of {$1.35\times10^{-11}$\,\flux} (i.e., about 40 times lower in comparison to the peak of the outburst), the source was also observed with the \nustar\ observatory in the broad energy band \citep{2021ATel14604....1T}. It was shown that broadband energy spectrum can be fitted with an absorbed cutoff power-law model with photon index of 1.25 and folding energy of 27\,keV. The absorption column density and blackbody component could not be constrained with the available data.
\maxi demonstrates, therefore, clear transient behaviour and was observed in a very wide range of luminosities. 

Both timing and spectral properties of accreting XRPs differ significantly at low and high mass accretion rates shedding light on different physical aspects of emission production and the interaction with matter in the presence of extreme magnetic fields. Here we focus, therefore, on reporting results of the detailed study of temporal and spectral properties of \maxi using the data from various X-ray instruments in the broad energy band at mass accretion rates differ by a factor of $\sim$30.

\begin{table}
        \caption{Observations of \maxi used in the current work. }
        \label{tab:log}
        \centering
        \begin{tabular}{lccc}
                \hline
                ObsID & $T_{\rm start}$ & $T_{\rm stop}$ & Exposure \\
                      &  MJD         &   MJD       &  ks      \\
                \hline
\multicolumn{4}{c}{{\it NuSTAR} observations}\\
                90701315002 & 59337.12 & 59337.59 & 23.3 \\
                90701320002 & 59375.63 & 59376.76 & 48.0 \\
                \hline
\multicolumn{4}{c}{SRG/ART-XC observations}\\
                11001000100 & 59001.89 & 59002.72  & 0.05 \\
                12000600100 & 59188.75 & 59189.59 & 0.05 \\
                12110050001 & 59357.64 & 59357.71 & 5.9 \\
                13000600100 & 59370.93 & 59371.59 & 0.02 \\
                \hline
\multicolumn{4}{c}{NICER observations}\\
4202120101 & 59320.82 & 59320.89 & 0.90 \\
4202120102 & 59321.02 & 59321.99 & 6.58 \\
4202120103 & 59322.24 & 59322.96 & 5.06 \\
4202120104 & 59323.02 & 59323.61 & 4.58 \\
4202120105 & 59324.51 & 59324.96 & 1.99 \\
4202120106 & 59325.02 & 59325.16 & 1.47 \\
4202120107 & 59326.45 & 59326.51 & 0.66 \\
4202120108 & 59327.22 & 59327.55 & 0.88 \\
4202120110 & 59329.42 & 59330.00 & 1.54 \\
4202120111 & 59330.06 & 59330.97 & 1.92 \\
4202120112 & 59331.09 & 59331.94 & 2.30 \\
4202120113 & 59331.99 & 59332.13 & 1.23 \\
4202120115 & 59335.93 & 59335.94 & 0.89 \\
\hline
\multicolumn{4}{c}{{\it Swift} observations}\\
00014281001 (wt) & 59320.97 & 59321.11 & 1.98 \\
00014281002 (wt) & 59327.35 & 59327.69 & 1.00 \\
00014281003 (wt) & 59329.74 & 59330.62 & 2.32 \\
00014281004 (wt) & 59331.08 & 59331.29 & 0.65 \\
00014281005 (wt) & 59333.72 & 59333.80 & 0.68 \\
00014281006 (pc) & 59378.46 & 59378.47 & 0.89 \\
00014281007 (pc) & 59384.97 & 59384.98 & 0.95 \\
00014281008 (pc) & 59387.15 & 59387.23 & 0.87 \\
00014281009 (pc) & 59390.01 & 59390.02 & 0.92 \\
00014281010 (pc) & 59393.41 & 59394.00 & 1.53 \\
00014281011 (pc) & 59396.19 & 59396.20 & 0.97 \\
00014281012 (pc) & 59398.70 & 59398.91 & 1.52 \\
00014281014 (pc) & 59405.27 & 59405.28 & 0.95 \\
00014281015 (pc) & 59408.34 & 59408.35 & 0.28 \\
00014281016 (pc) & 59411.17 & 59411.18 & 0.87 \\
00014281017 (pc) & 59414.09 & 59414.10 & 0.77 \\
00014281018 (pc) & 59417.01 & 59417.02 & 0.86 \\
00014281020 (pc) & 59423.06 & 59423.07 & 0.80 \\
00014281021 (pc) & 59426.92 & 59426.92 & 0.24 \\
00014281022 (pc) & 59429.84 & 59429.85 & 0.89 \\
00014281023 (pc) & 59432.96 & 59432.97 & 0.84 \\
                \hline
            
        \end{tabular}
\end{table}

\section{Data analysis}
\label{sec:data}
As already mentioned above, \maxi was observed with different X-ray and optical facilities since the discovery. Below we briefly describe the reduction procedures for each of the data-sets used in the current work. The list of the observations can be found in Table~\ref{tab:log}.

\subsection{\textit{NuSTAR}}

The main instrument for our study is the \textit{NuSTAR} observatory hosting two identical Focal Plane Modules~A and~B  \citep[FPMA and FPMB;][]{2013ApJ...770..103H}  covering the broad energy range of 3--79 keV. \nustar observed \maxi twice: first time on May 3, 2021 (ObsID 90701315002), about two weeks after the discovery of the source and thus already closer to the end of an outburst at a comparatively low luminosity \citep{2021ATel14604....1T}. The second observation was performed on June 10, 2021 (ObsID 90701320002) following trigger by the SRG/ART-XC telescope which discovered the resumed activity of \maxi, so the flux in the second observation is by a factor of 30  higher in comparison to the first \nustar pointing. 

The data for both observations were reduced following the standard procedures described in the {\it NuSTAR} user guide, and using the standard {\it NuSTAR} Data Analysis Software {\sc nustardas} v1.9.6 with the CALDB version 20210524.
The source and background spectra and light curves were extracted from circular regions with radii of 50\arcsec\ and 150\arcsec\ in the low state (observation 90701315002), and 60\arcsec\ and 150\arcsec\ in the bright state (observation 90701320002), respectively. The background in both cases was extracted from a source-free region in the corner of the field of view. The scientific products were obtained using the {\sc nuproducts} routine with the default parameters. Energy spectra were optimally rebinned using the prescription from \cite{2016A&A...587A.151K}. The lightcurves were background and baricentrically corrected. In order to increase statistics we added the lightcurves from both \nustar modules (FPMA and FPMB).

\subsection{SRG/ART-XC}

The {\it Spectrum Roentgen Gamma} (SRG) observatory \citep{sunyaev21} hosts two X-ray telescopes: the Mikhail Pavlinsky ART-XC operating in the 4--30\,keV energy range \citep{2021A&A...650A..42P} and \ero operating in the 0.2--10\,keV \citep{2021A&A...647A...1P}. This mission aims at the most sensitive all-sky survey in the X-ray band. 

SRG/ART-XC observed \maxi both during the surveys (on June 2, 2020, Dec 6, 2020, June 6, 2021) and during the technical stop (parking position) before the orbit correction on May 23, 2021. In the survey mode each source on the sky is observed once in half a year, with an exposure time on the order of several tens of seconds. In the low intensity state such an exposure allowed us to only detect the source and roughly estimate its flux. The exposure collected during the parking position is much longer (5.9\,ks). This observation allowed us to measure the source spectrum and determine its flux very accurately. During the third all sky survey scan, on June 6, 2021, the source was also bright enough to reconstruct its spectrum with ART-XC. As was mentioned above this observation was a trigger for the follow-up observation with \nustar. 

The ART-XC telescope consists of seven identical grazing incidence focusing modules with the total effective area of ${\sim}450\,\mathrm{cm}^2$ at 6\,keV, energy resolution of 1.4\,keV at 6\,keV, angular resolution of ${\sim}50$\arcsec\ and timing resolution of 23\,$\mu$s. ART-XC data were processed with the analysis software {\sc artproducts} v0.9 and with ART-XC CalDB version 20200401. A description of the ART-XC telescope and the software can be found in \cite{2021A&A...650A..42P}. All energy spectra were rebinned to have at least 1\,count per energy channel.

\subsection{NICER}

Soon after discovery of the source, the {\it Neutron star Interior Composition Explorer} \citep[NICER;][]{2012SPIE.8443E..13G} started its monitoring with about one day cadence. The monitoring continued for almost 40\,days and covered the whole outburst and some fraction of the quiescent period before the next outburst. In our work we use the NICER data collected until \maxi entered low-luminosity state when its flux became background dominated, i.e., above $F_{\rm 0.5-10\,keV}\sim10^{-11}$~\flux\ (or before MJD 59335). 

The raw NICER data were reduced using the {\sc nicerdas} software version 7a with default filtering criteria applied. The background was estimated using the tool {\sc nibackgen3C50}\footnote{\url{https://heasarc.gsfc.nasa.gov/docs/nicer/tools/nicer_bkg_est_tools.html}} (Remillard et al., in prep.) with the default parameters. Each spectrum was binned to have at least 1 count per energy channel.

\subsection{\textit{Swift}}

To study the long-term behaviour of the source including in the quiescent state, simultaneously with our \nustar observation in the bright state, we triggered a two-months monitoring of \maxi with the XRT telescope \citep{2005SSRv..120..165B} on-board the \textit{Neil Gehrels Swift Observatory} \citep{2004ApJ...611.1005G} with cadence of 3 days. Depending on the source brightness the data were collected either in the photon counting (PC) or windowed timing (WT) mode. The spectra for each observation were extracted using the online tools \citep{2009MNRAS.397.1177E} provided by the UK Swift Science Data Centre\footnote{\url{http://www.swift.ac.uk/user_objects/}} and binned to have at least 1\,count per energy channel.

\section{Results}
\begin{figure}
\centering
\includegraphics[width=0.97\columnwidth]{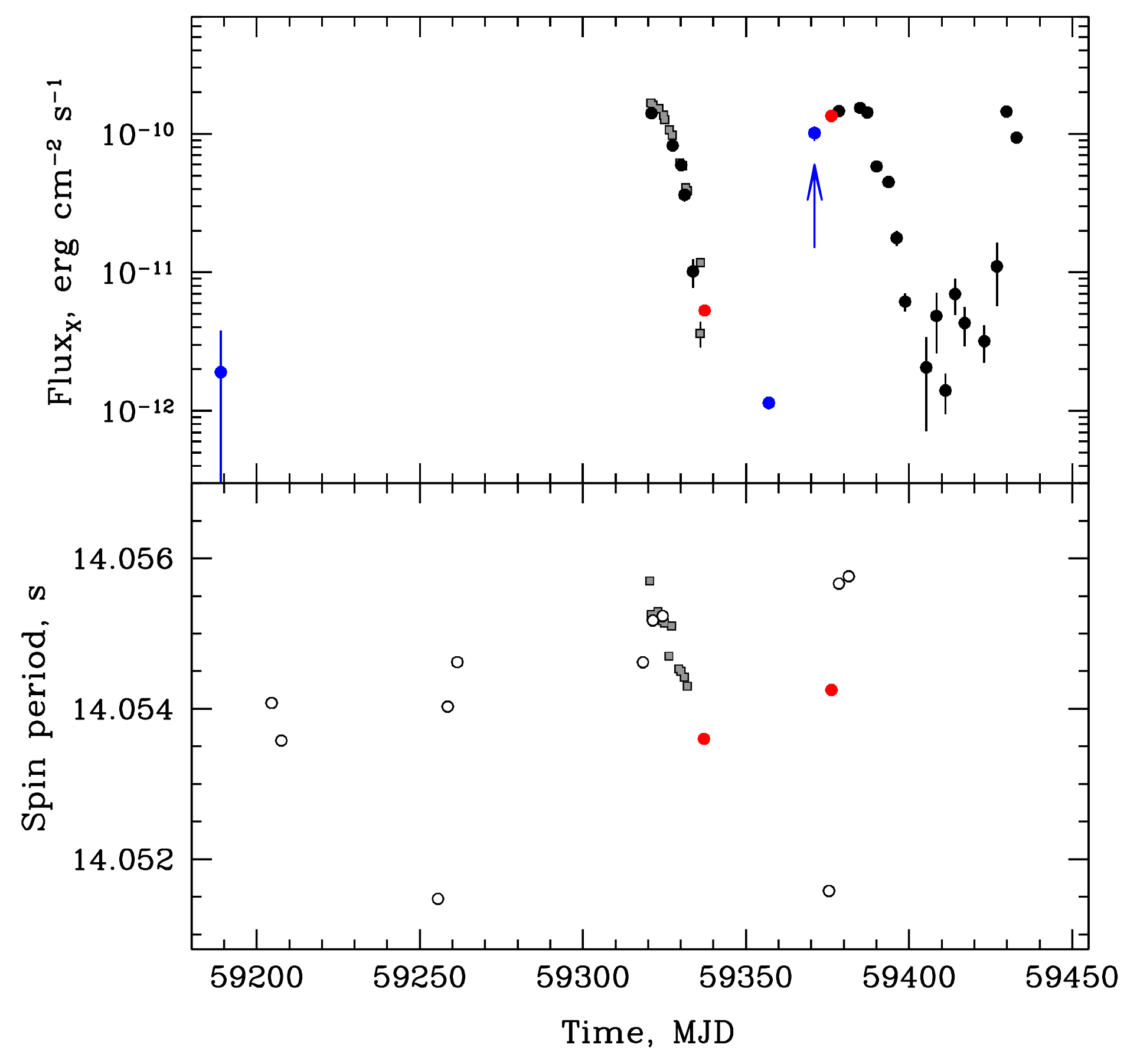}
\caption{{\it Top}:  The light curve of \maxi from the all available data. An observed flux is given in the 0.5-10 keV energy band for {\it Swift}/XRT (black) and NICER (gray), and in the 4-12 keV for SRG/ART-XC (blue) and {\it NuSTAR} (red). {\it Bottom}: Baricentrically corrected spin period of the pulsar obtained from {\it Fermi}/GBM (open circles), NICER (gray) and {\it NuSTAR} (red).
 }
 \label{fig:lc}
\end{figure}

Using all available observations described above we are able to study temporal and spectral properties of \maxi at different time scales and mass accretion rates. The overall light curve of the source covering about four months of observations is presented in Fig.~\ref{fig:lc}. In addition we discuss separately results obtained in two deep observations with \nustar which allow also detailed spectral analysis in broad energy band. 

\subsection{Timing analysis}
\maxi demonstrates strong flux variability at different time scales. As was shown by \cite{2021ATel14730....1C}, on the long time scale the source exhibits transient behaviour with period of around 57 days. The authors suggested that this activity is connected to periastron passages of the NS in an eccentric binary system. On short time scales flux from \maxi pulsates with the period of $\sim14$~s \citep{2021ATel14559....1R}. Here we discuss variations of spin frequency and pulse profiles of the source throughout the outburst.

\subsubsection{Spin frequency changes and binary orbit}
 Using the background subtracted and baricentrically corrected light curves, we measured spin period of pulsar in all available NICER and \nustar observations applying standard epoch folding technique \citep[{\sc efseach} tool from the {\sc ftools} package; ][]{1987A&A...180..275L}. Uncertainty on the spin period was estimated as described by \cite{2013AstL...39..375B}.
In addition, measurements provided by the {\it Fermi}/GBM XRPs monitoring project\footnote{\url{https://gammaray.nsstc.nasa.gov/gbm/science/pulsars/lightcurves/maxij0903.html}} were also used. Low count statistics   did not allow us to detect pulsations in the SRG/ART-XC data.
Because no orbital parameters except for the tentatively identified 57-days orbital period are known, no binary correction was applied to the light curves for initial timing analysis. The resulting dependence of apparent spin period of \maxi on time is shown in the bottom panel of Fig.~\ref{fig:lc}.

As can be seen from Fig.~\ref{fig:lc}, both the observed spin frequency and flux of \maxi exhibit $\sim57$\,d periodic variations likely associated with orbital modulation. Indeed, very strong apparent spin-down just before the outburst and gradual spin-up after the peak of the outburst is the opposite to what could be expected from an accreting neutron star under influence of accretion torque, and can only be explained by Doppler effect associated with orbital motion. On top of that a possible long-term spin-down trend is observed as was already mentioned by \cite{2021ATel14730....1C}.

To recover intrinsic spin evolution of the pulsar and orbital ephemerides we conducted, therefore, modeling of the observed spin frequency changes. To model observed spin history we used approach similar to that used by \cite{2016A&A...589A..72D}. However, considering that  only few measurements of spin frequency are available, and that those only cover limited orbital phase interval, we adopted a simplified approach. In particular, we described intrinsic spin history as a piecewise linear function characterized by steady spin-up/down during and between the outbursts, respectively. The frequency change rates were considered as free parameters in both cases but were assumed to equal for all burst/quiescence intervals. 
The reference epoch for spin-up episodes during the outbursts was assumed to coincide with periastron passage, and quiescent periods were assumed to start fifteen days later (which corresponds to the approximate duration of the bright phase of the outburst where spin-up can be expected) as the reference epoch for inter-burst periods. To avoid  jumps in the assumed spin frequency the resulting piece-wise function was in addition smoothed with fifteen days long Hann window. This actually mimics quite well the expected changes of spin frequency due to accretion torque in a series of short outbursts.

\begin{figure}
    \centering
    \includegraphics[width=\columnwidth]{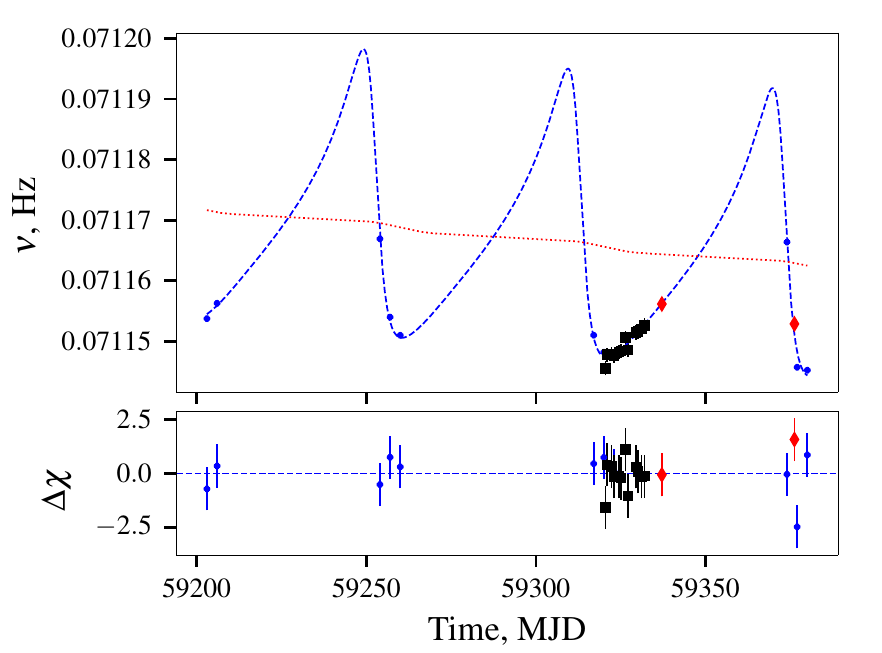}
    \caption{Spin history of the source as observed by \textit{Fermi}/GBM (blue dots), NICER (black squares) and \textit{NuSTAR} (red diamonds). The final model including the intrinsic spin frequency variations (red) and those due to the orbital motion (blue) together with the best-fit residuals (lower panel) are also shown.}
    \label{fig:orbit}
\end{figure}

\begin{table}
    \caption{Best-fit orbital parameters assuming constant spin-up during outbursts and spin-down between the outbursts. Uncertainties are reported at $1\sigma$ confidence level.}
    \label{tab:orbit}
    \centering
    \begin{tabular}{l|l}
    \hline
     Parameter & Value  \\
     \hline
     $P_{\rm orb}$, d & 60.3(1) \\ 
     $a\sin{i}$, lt. s & 225(6) \\
     $e$ & 0.56(4) \\
     $\omega$, deg & 247(17) \\
     $T_{\rm PA}$, MJD & 59192(3) \\
     $\nu_0$,10$^{-2}$ Hz & 7.1173(2)\\
     $\dot{\nu}_{\rm o}$, $10^{-7}$\,Hz\,d$^{-1}$ & -2(5)\\
     $\dot{\nu}_{\rm q}$, $10^{-8}$\,Hz\,d$^{-1}$ & -2(16)\\
     \hline
     $\chi^2$/dof & 16.78/17 \\ 
     \hline
    \end{tabular}
\end{table}

\begin{figure}
\includegraphics[width=0.97\columnwidth]{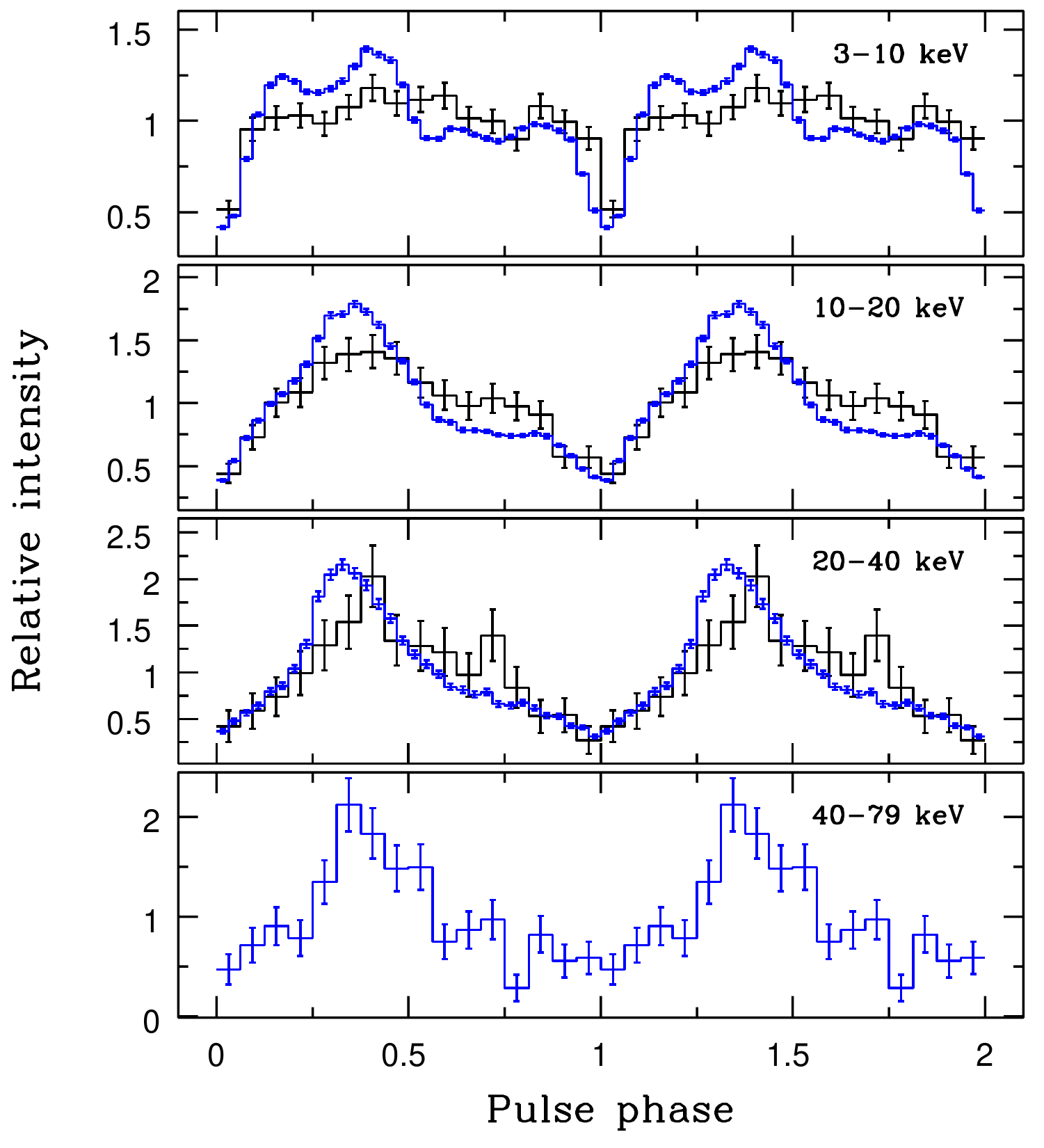}
\caption{Pulse profiles of \maxi obtained from the \nustar data in different energy bands. Data for both observations are presented, in black for the low state (ObsID 90701315002) and blue for the high state (ObsID 90701320002). Zero phase was chosen arbitrary to match the minimum in the profile. 
}
 \label{fig:pprof}
\end{figure}

We emphasize that the model described above is just a rough approximation of actual intrinsic spin frequency changes. The available data, however, is insufficient to model it more accurately based on the observed flux like was done, i.e. in \cite{2017A&A...605A..39T, 2018A&A...613A..19D}. On the other hand, also the accuracy of spin frequency measurements and contribution of intrinsic spin variations is lower for \maxi, so accurate description of intrinsic frequency variations is less important. We verified, for instance, that the derived orbital parameters are almost unaffected if one assumes steady spin-down instead of the more complex model adopted above. This is also confirmed by the fact that best-fit spin-up rate for outbursts is both consistent with zero and expected spin-up rate for a neutron star with standard parameters accreting at observed luminosity of $\sim4.5\times10^{36}$\,erg\,s$^{-1}$. Indeed, an upper limit on the expected spin-up rate might be derived assuming that pulsar is close to co-rotation at $\dot{\nu}\le\dot{M}\sqrt{GMR_c}$, where $R_c=(GM/(2\pi\nu)^2)^{1/3}\sim10^{9}$\,cm is the co-rotation radius. The resulting value of $2.6\times10^{-7}$\,Hz day$^{-1}$ is consistent with our best-fit value, which is also consistent with zero, so we can conclude that quality of available frequency measurements is insufficient to study intrinsic variations of the spin frequency in detail and that the observed spin evolution is dominated by orbital motion.  The derived orbital parameters shall, therefore, be only weakly sensitive to assumptions regarding the intrinsic spin variations. The best-fit results are presented in Fig.~\ref{fig:orbit} and Table~\ref{tab:orbit}. Note that the reported uncertainties do not account for systematic effects mentioned above and only reflect statistical uncertainties. The combination of the orbital period and eccentricity obtained for \maxi is typical for the Galactic Be/XRPs \citep[e.g.][]{2011MNRAS.416.1556T}.

\subsubsection{Energy dependence of the pulse profiles}

\begin{figure}
\includegraphics[width=0.97\columnwidth]{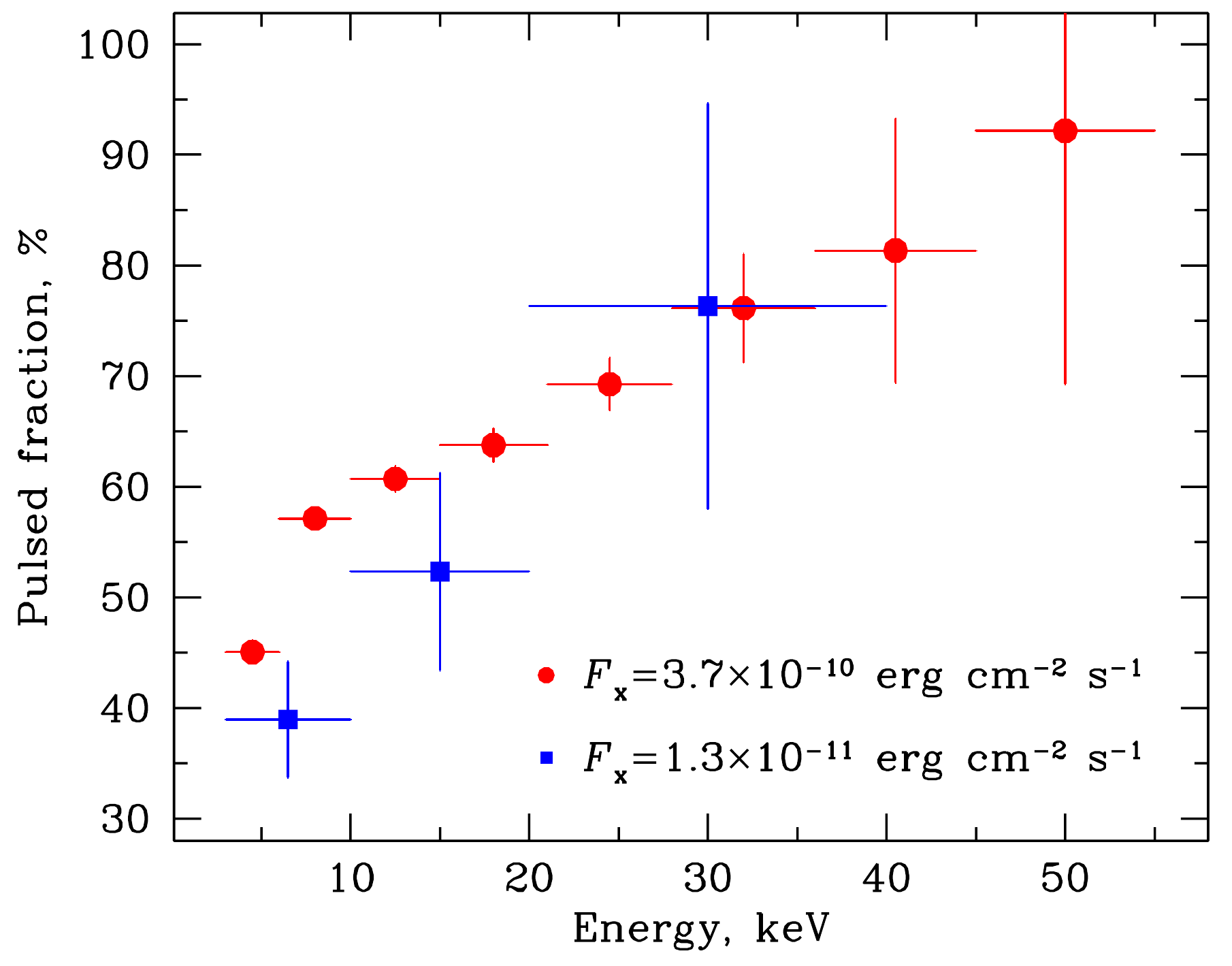}
\caption{Dependence of the pulsed fraction of \maxi on energy band based on the \nustar data in different states of the source.
}
 \label{fig:ppfrac}
\end{figure}

Sufficiently high count statistics in the \nustar data allowed us to study pulse profile shape as a function of energy range and the source luminosity. 
As can be seen from Fig.~\ref{fig:pprof} the source profile in general is single peaked and depends on energy only slightly. Particularly, at energies below 10 keV it consists of four sub-peaks with increasing domination of one of them at higher energies. This picture is characteristic for both high and low states of the source. At the highest energies (above 40 keV) pulsations are detected only in the brightest observation due to better counting statistics. The pulsed fraction (defined as $(F_{\rm max} - F_{\rm min})/(F_{\rm max} + F_{\rm min})$, where $F_{\rm max}$ and $F_{\rm min}$ are maximum and minimum fluxes in the pulse profile, respectively) demonstrates an increase from 40\% to roughly 80\% for both states (see Fig.~\ref{fig:ppfrac}). Such behaviour is typical for the majority of XRPs \citep{2009AstL...35..433L}.

\subsection{Spectral analysis}
\label{sec:spec}

Spectral properties of \maxi in the low state were studied based on the \nustar data (ObsID 90701315002) by \cite{2021ATel14604....1T}. The authors showed that the spectrum can be fitted with the model consisting of a simple absorbed cutoff power-law ({\sc cutoffpl} in {\sc xspec}) with a photon index of $1.25^{+0.19}_{-0.14}$, $E_{\rm fold}=27^{+18}_{-7}$~keV and hydrogen column density fixed at $N_{\rm H} = 4.1\times10^{21}$~cm$^{-2}$.
To facilitate comparison of the spectra of \maxi in different intensity states, we reanalysed all available data uniformly. To expand our analysis to the lower energies we jointly fitted the \nustar spectra with the {\it Swift}/XRT data collected several days apart from the \nustar observations. Particularly, \nustar observation 90701315002 was analysed together with XRT observation 00014281005, and \nustar observation 90701320002 with the XRT observation 00014281006 (see Tab.~\ref{tab:log}). The data from {\it Swift}/XRT and {\it NuSTAR} were used in the 0.3--10 keV and 3--79~keV bands, respectively. Both broadband spectra were also supplemented with the SRG/ART-XC data in the 4-20 keV energy band (ObsIDs 12110050001 and 13000600100, respectively). To fit the data we used the {\sc xspec} package \citep{Arn96} applying W-statistic\footnote{\url{https://heasarc.gsfc.nasa.gov/xanadu/xspec/manual/XSappendixStatistics.html}} \citep{1979ApJ...230..274W} appropriate for modeling of the background-subtracted spectra in low count regime.

\begin{figure}
\includegraphics[width=0.97\columnwidth]{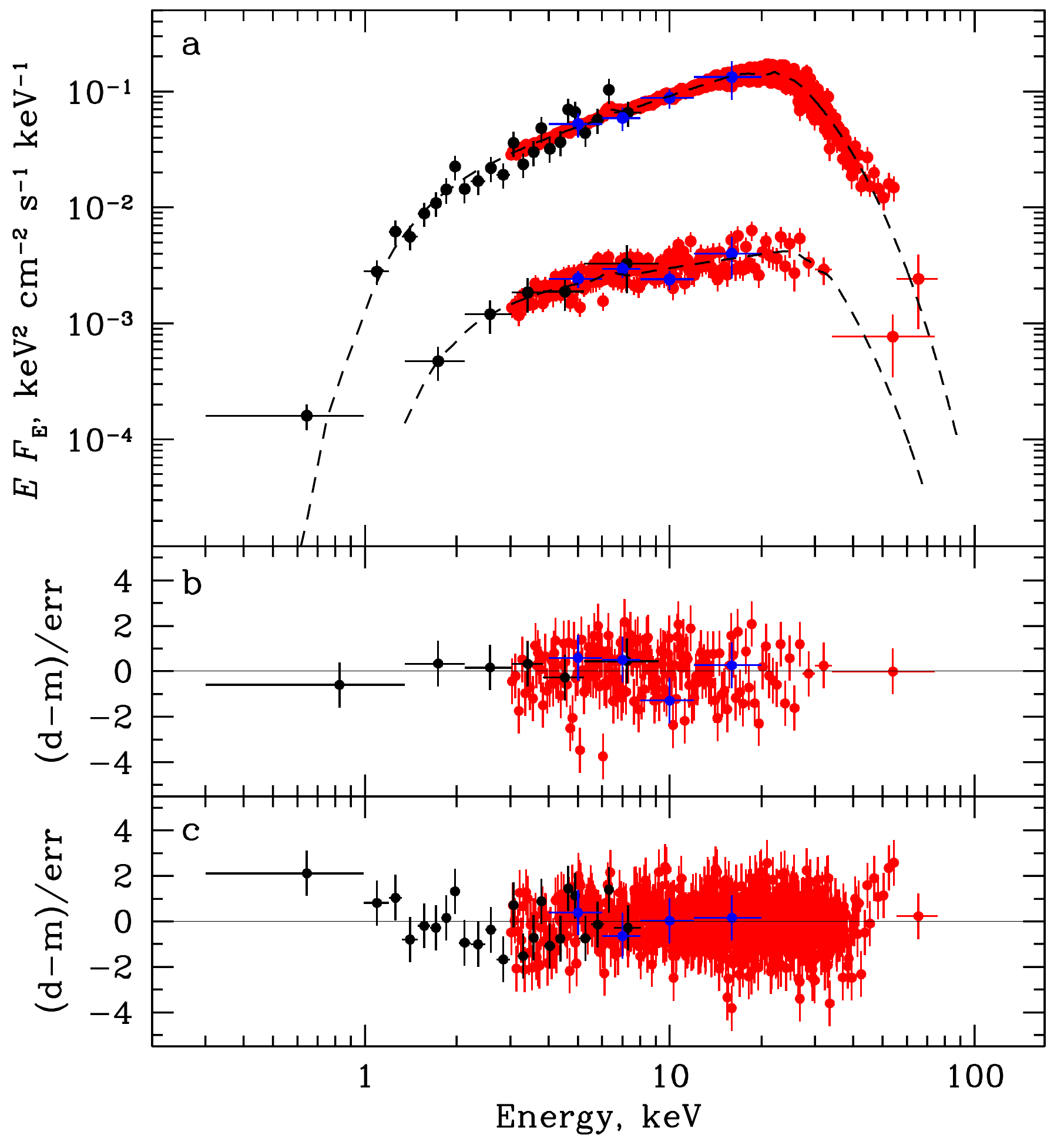}
\caption{Unfolded energy spectra of \maxi in different states with luminosity varying by a factor of $\sim30$ (panel a). 
The black, red and blue points correspond to the {\it Swift}/XRT, {\it NuSTAR} and SRG/ART-XC data, respectively. Data from two {\it NuSTAR} modules are combined for the illustrative purposes only.
The dashed lines represent the best-fitting models listed in Table~\ref{tab:spec}. 
The corresponding residuals for the low and high luminosity states of \maxi are presented at the panels b and c, respectively. 
}
 \label{fig:spec}
\end{figure}

\begin{table}
        \begin{center}
        \caption{Best-fitting results of the broad-band spectra approximation for  \maxi in different states obtained with {\it NuSTAR}, {\it Swift}/XRT and SRG/ART-XC. }
        \label{tab:spec}
        \begin{tabular}{lcc}
\hline
  Parameter  & Low state & High state \\
\hline
$N_{\rm H}$, $10^{22}$ cm$^{-2}$  & 3.3$\pm$1.1 & 1.3$\pm$0.2 \\[1ex]
Phot. index                     & 1.64$\pm$0.05 & 1.14$\pm$0.01 \\[1ex]
$E_{\rm cut}$, keV.             &   28.4$^{+4.6}_{-5.1}$ & 22.2$\pm$0.3 \\[1ex]
$E_{\rm fold}$, keV               &   7.8$^{+6.0}_{-2.3}$ & 7.5$\pm$0.2 \\[1ex]
$E_{\rm Iron}$, keV              &  6.29$^{a}$ & 6.29$\pm$0.04 \\[1ex]
$\sigma_{\rm Iron}$, keV         &  0.33$^{a}$ & 0.33$\pm$0.08 \\[1ex]
$Norm_{\rm Iron}$                &  $(5.2\pm3.8)\times10^{-6}$ & $(1.5\pm0.2)\times10^{-4}$ \\[1ex]
$C_{\rm XRT}$                   & 1.6$\pm$0.3 & 1.05$\pm$0.06 \\[1ex]
$C_{\rm FPMA}$                   &  1  &  1  \\[1ex]
$C_{\rm FPMB}$                   & 1.03$\pm$0.03 & 1.047$\pm$0.004 \\[1ex]
$C_{\rm ART-XC}$                 & 0.29$\pm$0.02 & 0.98$\pm$0.05 \\[1ex]
$F_{\rm X}$ $^{b}$, $10^{-10}$ \flux   & 0.129$\pm$0.0.006 & 3.73$\pm$0.02 \\[1ex]
C-stat (d.o.f.)                & 404.2 (409) & 756.9 (703) \\[1ex]
\hline
        \end{tabular}
	\begin{tablenotes}
        \item $^{a}$ Iron line parameters were fixed at values derived from the spectrum obtained in the bright state.  
        \item $^{b}$ The observed fluxes given in the 0.5--100 keV energy range.  
        \end{tablenotes}
\end{center}
\end{table}

In order to get acceptable description of the source spectrum we used several phenomenological models broadly applied to fit continua of XRPs. We started with {\sc cutoffpl} modified with a photoelectric absorption model \citep[{\sc tbabs} in {\sc xspec}, with abundances from ][]{Wilms2000} and fluorescent emission iron line in the form of additive Gaussian component ({\sc gau} in {\sc xspec}). Although this model fits the spectrum of \maxi in the low state, it is unable to reproduce its spectrum in the high state resulting in the C-stat(d.o.f.)=3345.5(707). Comptonization model ({\sc comptt} in {\sc xspec}) also produces strong residuals around 10 and 40 keV with resulting C-stat(d.o.f.)=1313.8(707). 

Finally we were able to get perfect fit using an absorbed power-law model modified with the high-energy exponential cutoff and iron line ({\sc tbabs$\times$(po$\times$highecut+gau)}; see Fig.~\ref{fig:spec}). To get rid of an artificial absorption-like residuals around the cutoff energy $E_{\rm cut}$ inherently present in this continuum model due to discontinuity at the cutoff, we also introduced a narrow and shallow ($\sigma=0.1 E_{\rm cut}$~keV, $\tau=0.1$) negative Gaussian with its central energy tied to the cutoff energy \citep[see e.g.][]{2002ApJ...580..394C}.
In order to compare spectra of \maxi in two states, the same model was  applied to both broadband data-sets. The resulting parameters of the best-fit model are presented in Table~\ref{tab:spec}. Parameters of the spectral model are typical for the accreting XRPs \citep[e.g.][]{2002ApJ...580..394C,2005AstL...31..729F}. The absorption value revealed by the fit is just slightly higher in comparison to the Galactic mean value in the direction to the source of $0.8\times10^{22}$ cm$^{-2}$ \citep{2013MNRAS.431..394W} that is typical for the Be/XRPs. Also as can be seen from the residuals in Fig.~\ref{fig:spec} no significant deviations which can be interpreted as, e.g. a cyclotron resonant scattering feature, can be found in the spectrum.

\begin{figure}
\includegraphics[width=0.97\columnwidth]{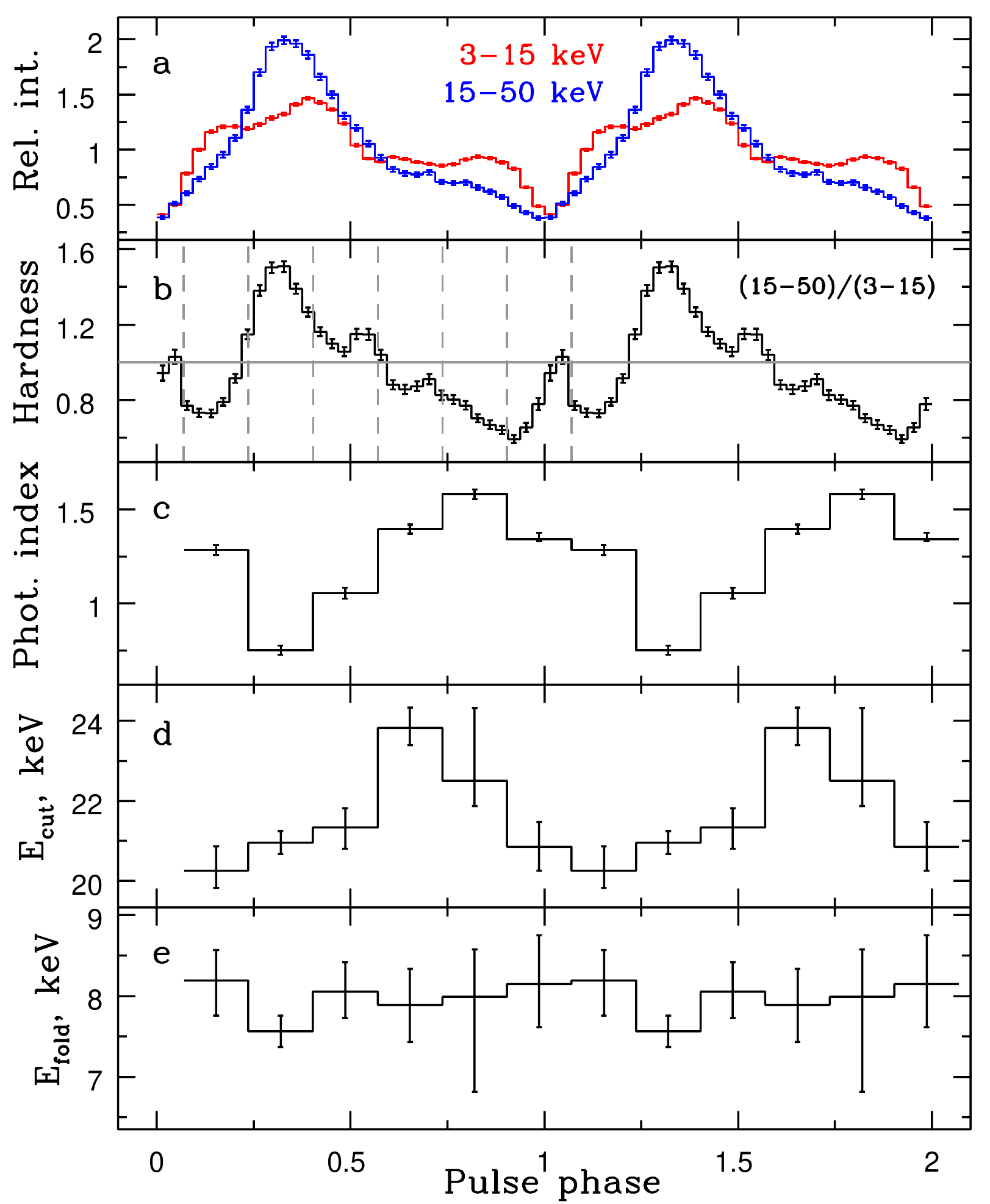}
\caption{Results of the pulse phase-resolved spectral analysis of \maxi using the \nustar data in the bright state. To illustrate spectral variations over the pulse phase, the normalised pulse profiles in 3-15 keV and 15-50 keV bands and their ratio are shown in panel a and b, respectively. Panels c, d and e demonstrate behaviour of the photon index, cut-off and folding energies, respectively. 
}
 \label{fig:phres}
\end{figure}

To verify the conclusion about an absence of additional spectral features appearing only at some specific phases of rotation of the NS, as observed in some XRPs \citep[see e.g.][]{2019ApJ...883L..11M,2021ApJ...915L..27M}, we produced the phase-resolved energy spectra of \maxi using the \nustar data in the bright state. Particularly, 6 evenly distributed phase bins were chosen as shown in Fig.~\ref{fig:phres}. All spectra were fitted with our best-fit model with iron line energy and width fixed at the values obtained from the averaged spectrum. As can be seen from Fig.~\ref{fig:phres}b hardness of emission is significantly variable over the pulse mainly due to changes of the photon index. We did not find, however, any evidence for any additional emission or absorption features at any phase, nor for significant variations of the $N_{\rm H}$ value.

\section{Discussion}
\label{sec:discus}

NSs (magnetars and XRPs) are considered as the strongest magnets in the Universe. The only direct way to measure the strength of the magnetic field in such systems is to detect a cyclotron resonant scattering feature in the energy spectrum \citep[see e.g.][]{2019A&A...622A..61S}. As was discussed in Sec.~\ref{sec:spec}, we did not detect any absorption features in the \maxi spectrum. However, the source demonstrates strong flux variability on different time scales. Behaviour of an XRP when it transits from a high accretion rate regime to the quiescence can be used to make some conclusions about the NS magnetic field.

\subsection{Magnetic field of the NS}
\label{sec:bfield}

It was shown theoretically \citep{1975A&A....39..185I} and found observationally for several XRPs \citep{1997ApJ...482L.163C,2016MNRAS.457.1101T,2016A&A...593A..16T, 2017ApJ...834..209L} that in the case of fast rotating NS with strong magnetic field the so called propeller effect is expected to stop the accretion abruptly below certain mass accretion rate. This effect takes place when the magnetospheric and corotation radii are about equal. The corresponding luminosity can be estimated as  \citep[see, e.g.][]{2002ApJ...580..389C}:
\be\label{eq_prop} 
L_{\rm prop}
\simeq \frac{GM\dot{M}}{R} \simeq 4 \times 10^{37} k^{7/2}
B_{12}^2 P^{-7/3} M_{1.4}^{-2/3} R_6^5 \,\textrm{erg s$^{-1}$} , 
\ee
where $\dot{M}$ is the mass-accretion rate,  $B_{12}$ is the NS magnetic field in units $10^{12}$~G, $P$ is the NS spin period in seconds,  $M_{1.4}$ is the NS mass in units $1.4M_{\odot}$  and $R_6$ is the radius in $10^6$~cm.  
The factor $k$ accounts for the details of interaction of the accretion flow with the magnetosphere, relating its size to the Alfv\'en radius, that is, $k=R_{\rm m}/R_{\rm A}$. In the case of disc accretion, it is usually assumed to be $k=0.5$ \citep{GL1978}.

As can be seen from Eq.~(\ref{eq_prop}), the critical luminosity $L_{\rm prop}$ is a strong function of the pulsar magnetic field and hence can be used to constrain its value. \maxi does not demonstrate sharp drops of the flux (see Fig.~\ref{fig:lc}) and has hard X-ray spectrum  in the quiescent state (with photon index $1.3\pm0.3$ measured from the XRT data averaged over MJD 59405-59423) with luminosity around $3\times10^{34}$~\lum\ (assuming distance of 10 kpc and bolometric correction factor of 2). Using this luminosity as an upper limit on the critical luminosity $L_{\rm prop}$ we can derive upper limit on the dipole component of the NS magnetic field as $2\times10^{12}$ G assuming standard mass of $1.4M_{\odot}$ and radius of $10^6$~cm. Adoption of 12 km radius of the NS would decrease its magnetic field roughly twice. These estimates point to a rather low magnetic field in \maxi.

At the same time, the observed stability of the source in the quiescence at around several $\times10^{34}$~\lum\  evidences the transition of the source to the state of a quasi-stable cold-disc accretion due to the reduced disc viscosity at low mass accretion rates much like in dwarf novae \citep{2017A&A...608A..17T}. 
Indeed, transitions to the cold-disc state have been already observed in several XRPs \citep[e.g.,][]{2018A&A...620L..13R,2019A&A...622A.198N,2019A&A...621A.134T,2020A&A...634A..89D} at luminosities around $\sim10^{34-35}$~\lum. The luminosity corresponding to the transition to this state can be calculated by the equating the inner radius of the accretion disc when temperature there reaches 6500~K to the magnetospheric radius of the NS and, hence, can be also used to estimate its magnetic field \citep{2017A&A...608A..17T}. Thus, using eq.~(12) from \cite{2017A&A...608A..17T} and transitional luminosity $L_{\rm cold}\sim3\times10^{34}$~\lum, we obtained magnetic field $\sim3\times10^{12}$ G similar to one obtained above based on the propeller effect.

\subsection{Spectral evolution}

Deep \nustar observations of \maxi in two states differ by a factor of $\sim30$ in flux allowed us to perform detailed spectral analysis and to compare spectral shape at very different mass accretion rates. The goal of that was to verify an assumption about a strong changes of the spectra in XRPs when their luminosity reaches $\sim10^{35}$~\lum\ as was recently found in several X-ray pulsars (X~Persei, \citealt{2012A&A...540L...1D}; GX~304$-$1, \citealt{2019MNRAS.483L.144T}; A~0535+262, \citealt{2019MNRAS.487L..30T}; GRO J1008$-$57, \citealt{2021ApJ...912...17L}; 2SXPS~J075542.5$-$293353, \citealt{2021A&A...647A.165D}; SRGA~J124404.1$-$632232/SRGU J124403.8$-$632231, \citealt{2021arXiv210614539D}). 
These spectral changes manifested in the transformation of the classical pulsar-like into a double-hump spectrum  were shown to be caused by the Comptonisation of cyclotron photons in the upper over-heated layer of the NS atmosphere \citep{2021MNRAS.503.5193M,2021A&A...651A..12S}. Since in the low state of \maxi it was observed by \nustar with luminosity of around $\sim10^{35}$~\lum, one can expect to see similar spectral changes in the source.

However, as can be seen from Fig.~\ref{fig:spec} and Tab.~\ref{tab:spec} the spectrum of \maxi does not exhibit any dramatic variations and appears to be consistent between the two \nustar observations. We note that the similarity of X-ray spectra in two luminosity states can be caused by relatively low ($\lesssim (2-3)\times10^{12}\,{\rm G}$) magnetic field strength at the surface of a NS. Indeed, the high-energy component observed in low states of some XRPs is due to accretion flow braking and subsequent Comptonisation of seed cyclotron photons in the atmosphere of a NS.
This component of X-ray spectra appears at around cyclotron energy.
In the case of relatively low magnetic field, the high-energy hump of the spectrum arises at lower energy and merges with the low-energy hump caused by thermal radiation of the atmosphere. 
Additionally, the X-ray radiation leaving the atmosphere of a NS experiences resonant Compton scattering by the infalling material in the accretion channel above the hot spots.
Because of a Doppler effect in accretion channel \citep{2015MNRAS.454.2714M}, the resonant scattering affects X-ray photons from a wide energy range: $0.6E_{\rm cyc}\lesssim E \lesssim E_{\rm cyc}$.
The photons scattered resonantly above the stellar surface experience bulk Comptonisation \citep{2007ApJ...654..435B} and contribute to the high-energy tail of the spectrum.
As a result, the resonant bulk Comptonisation together with merged humps in X-ray spectra largely causes the observed spectral shape in XRPs at low mass accretion rate. 
Therefore, we argue that observation of a classical pulsar-like spectrum in the low-luminosity state points to a relatively weak magnetic field of the NS, that in the case of \maxi is supported by our indirect constraints (see Sec.~\ref{sec:bfield}).

\section{Conclusion}
 In this work we presented results of the broadband spectral and timing analysis of emission from the recently discovered transient XRP \maxi, obtained with the {\it NuSTAR}, {\it Swift}/XRT and SRG/ART-XC instruments at two luminosity states differ by a factor of $\sim30$. The data revealed that the spectrum in both states can be described as a classical pulsar-like spectrum consisting of power-law with high-energy cutoff. Both phase-averaged and resolved spectra do not contain any absorption components which can be interpreted as a CRSF.

We argue that absence of the transition of the spectrum from typical cutoff power law spectrum to the two-component spectrum at low luminosity observed in other pulsars and, expected also in the case of \maxi in the low state, points to a relatively weak magnetic field of the NS below $(2-3)\times10^{12}$~G. This value is compatible with the constraints obtained from the propeller effect and transition of the source to the accretion from the cold disc.

Timing analysis of the \nustar\ data revealed only slight variations of a single-peaked pulse profile of the source as a function of energy band and mass accretion rate. In both intensity states the pulsed fraction demonstrates increasing from 40\% to roughly 80\%. We were also able to obtain the orbital solution for the binary system using data from the {\it Fermi}/GBM, NICER and {\it NuSTAR} instruments.

\begin{acknowledgements}
This work is based on observations with  Mikhail Pavlinsky \art\ telescope, hard X-ray instrument on board the \srg\  observatory. The \srg\ observatory was created by Roskosmos (the Lavochkin Association and its subcontractors) in the interests of the Russian Academy of Sciences represented by its Space Research Institute (IKI) in the framework of the Russian Federal Space Program, with the participation of Germany. The \art\ team thanks the Russian Space Agency, Russian Academy of Sciences and State Corporation Rosatom for the support of the  \art\ telescope design and development. The science data are downlinked via the Deep Space Network Antennae in Bear Lakes, Ussurijsk, and Baykonur, funded by Roskosmos. We grateful to the {\it NuSTAR} and {\it Swift} teams for approving the DDT observations of \maxi. This research has made use of data and software provided by the High Energy Astrophysics Science Archive Research Center (HEASARC), which is a service of the Astrophysics Science Division at NASA/GSFC and the High Energy Astrophysics Division of the Smithsonian Astrophysical Observatory. It also made use of data supplied by the UK Swift Science Data Centre at the University of Leicester. This work was supported by the grant 14.W03.31.0021 of the Ministry of Science and Higher Education of the Russian Federation. We also acknowledge the support from the Academy of Finland travel grants 331951,  324550 (SST), the V\"ais\"al\"a Foundation (SST),  the  Netherlands Organization for Scientific Research Veni Fellowship (AAM). 
\end{acknowledgements}

\bibliographystyle{aa}
\bibliography{allbib}

\end{document}